# Improving the Time Resolution of Large-Area $LaBr_3$:Ce Detectors with SiPM Array Readout

**Maurizio Bonesini [1,2,*], Roberto Bertoni [1], Andrea Abba [3], Francesco Caponio [3], Marco Prata [4] and Massimo Rossella [4]**

[1] Sezione INFN Milano Bicocca, Piazza Scienza 3, 20123 Milano, Italy
[2] Dipartimento di Fisica G. Occhialini, Universitá Milano Bicocca, Piazza Scienza 3, 20123 Milano, Italy
[3] Nuclear Instruments srl, via Lecco 3, 22045 Lambrugo, Italy
[4] Sezione INFN Pavia, Via A. Bassi 6, 27100 Pavia, Italy
* Correspondence: Maurizio.Bonesini@mib.infn.it

## 1. Introduction

Ce:$LaBr_3$ crystals have extensive applications in radiation imaging in medical physics [1,2], homeland security [3,4] and gamma-ray astronomy [5,6]. The adoption of a readout based on a SiPM or a SiPM array instead of a conventional photomultiplier (PMT) allows the realization of compact detectors and their use in strong external magnetic fields.

Many efforts have been made to optimize large-area detectors with SiPM readouts (area 1″ or more) to increase both the FWHM energy resolution [7,8] and the signal timing properties (risetime/falltime) [9]. FWHM energy resolutions around 3% or better were reached at the $Cs^{137}$ photopeak (661.7 keV) in [7] with 3″ Ce:$LaBr_3$ crystals, and signal risetime less than 10 ns were obtained in [9] with a small 3 mm × 3 mm × 5 mm crystal read by a single 3 × 3 $mm^2$ Hamamatsu S10362-33-050C SiPM. These results compare well with the best ones obtained with a PMT readout [10,11]. Unfortunately, until now, it is difficult to combine good timing properties (risetime/falltime) with a small FWHM energy resolution ( around 3% at the $^{137}$Cs photopeak) in large-area detectors with SiPM readouts. Our efforts aimed at obtaining this goal: initially with 1/2″ crystals and then with 1″ ones [12]. Our studies were pursued in the framework of the FAMU (**F**isica degli **A**tomi **Mu**onici) project [13–15] at Port 1 of the RIKEN-RAL muon facility [16], whose aim is the high-precision measurement of the hyperfine splitting (HFS) of the μp ground state and thus of the proton Zemach radius [17]. Similar experiments were also proposed at PSI [18] and JPARC [19].

FAMU may contribute to solving the so-called "proton radius puzzle", where a large discrepancy was found in the proton charge, as measured with impinging electrons or muons [20–22]. Even if new data with incoming electrons from the PRad collaboration [23] have shown now a good agreement with existing electron data, one needs to understand why there are still discrepancies with previous experiments. The innovative method introduced by FAMU [24,25] implies the detection of characteristic X-rays around 130 keV. In addition, to separate the "delayed signal" component from the prompt background, a fast response from the used detectors is needed, particularly short signal falltimes well below 300–400 ns.

## 2. Detectors' development

After preliminary studies with non-hygroscopic crystals, such as Pr:LuAg and Ce:GAGG [26,27], Ce:$LaBr_3$ crystals were chosen for their better energy resolution and faster decay time, notwithstanding their hygroscopicity. As the main aim was the detection of X-rays around 130 keV, a reduced crystal's thickness of 1/2″ was found to be sufficient [28], from an estimate based on tabulated X-ray attenuation coefficients [29] and a complete simulation based on the MNCP code [30]. The Ce:$LaBr_3$ crystal and the PCB on which the SiPM array is mounted are housed inside a 3D-printed ABS holder, as shown in the right panel of Figure 1 . Hamamatsu S14161-6050AS-04 1″ square SiPM arrays are used for the readout. With $6 \times 6$ mm$^2$ cells, they have an operating voltage $V_{op} \sim 41.1$ V, with a maximum PDE around 50% at ∼450 nm. Additional details on detectors' construction are reported in reference [28].

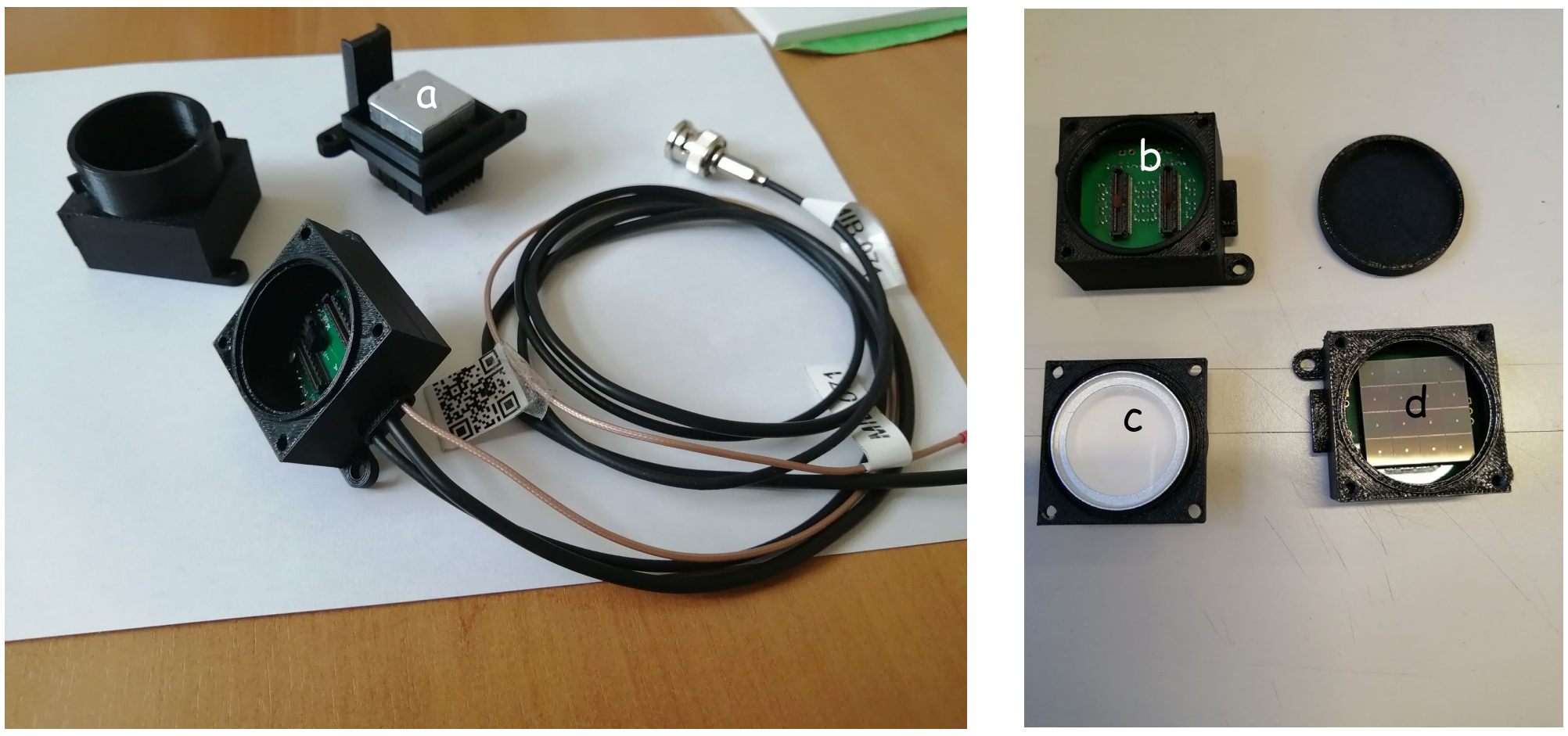

**Figure 1.** Components of a 1″ Ce:LaBr$_3$ detector. All are printed with a 3D printer. (**a**) Bottom closure, equipped with a power dissipator; (**b**) PCB seen from top: the two SAMTEC connectors for SiPM array mounting are shown ; (**c**) LaBr$_3$:Ce crystal inside the holder; (**d**) mounted S14161-6050-AS array with silicone window.

The sum of the signals from the 16 array's SiPM cells is then digitized with a CAEN V1730 FADC. The different cells may be powered using different "ganging" schemes that have a relevant influence on the signal pulse shape (especially the falltime). Below, results with different ganging schemes are shown: from standard parallel ganging to hybrid ganging and finally to the 4-1 innovative scheme developed by Nuclear Instruments.

As the breakdown voltage $V_{brk}$ of SiPM changes with temperature according to

$$V_{brk}(T) = V_{brk}(T_{ref}) \times (1 + \beta(T - T_{ref}))$$

with T$_{ref}$ reference temperature (typically 25 °C) and $\beta = \Delta V_{brk}/\Delta T$ temperature coefficient of the used SiPM (34 mV/C for Hamamatsu S14161), their operating voltage $V_{op} = V_{brk} + \Delta V$, where ΔV is the overvoltage, must be changed accordingly to keep a fixed gain and the same value of the PDE. As explained in reference [31], $\beta$ is independent of the temperature T. The temperature T is measured on the back side of the SiPM arrays via an Analog Devices TMP37 thermistor. This information is then used by a custom NIM module, based on CAEN A7585D electronic modules, to correct online the operating voltage (see references [32,33] for more details). As shown in Figure 2, the effect on the detector response (pulse height (P.H.) of the Cs$^{137}$ photopeak in a.u.) between 10 °C and 35 °C is reduced from 40 % to 10 % for 1″ detectors. Measurements were performed inside a IPV30 Memmert climatic chamber with a ±0.1°C temperature control.

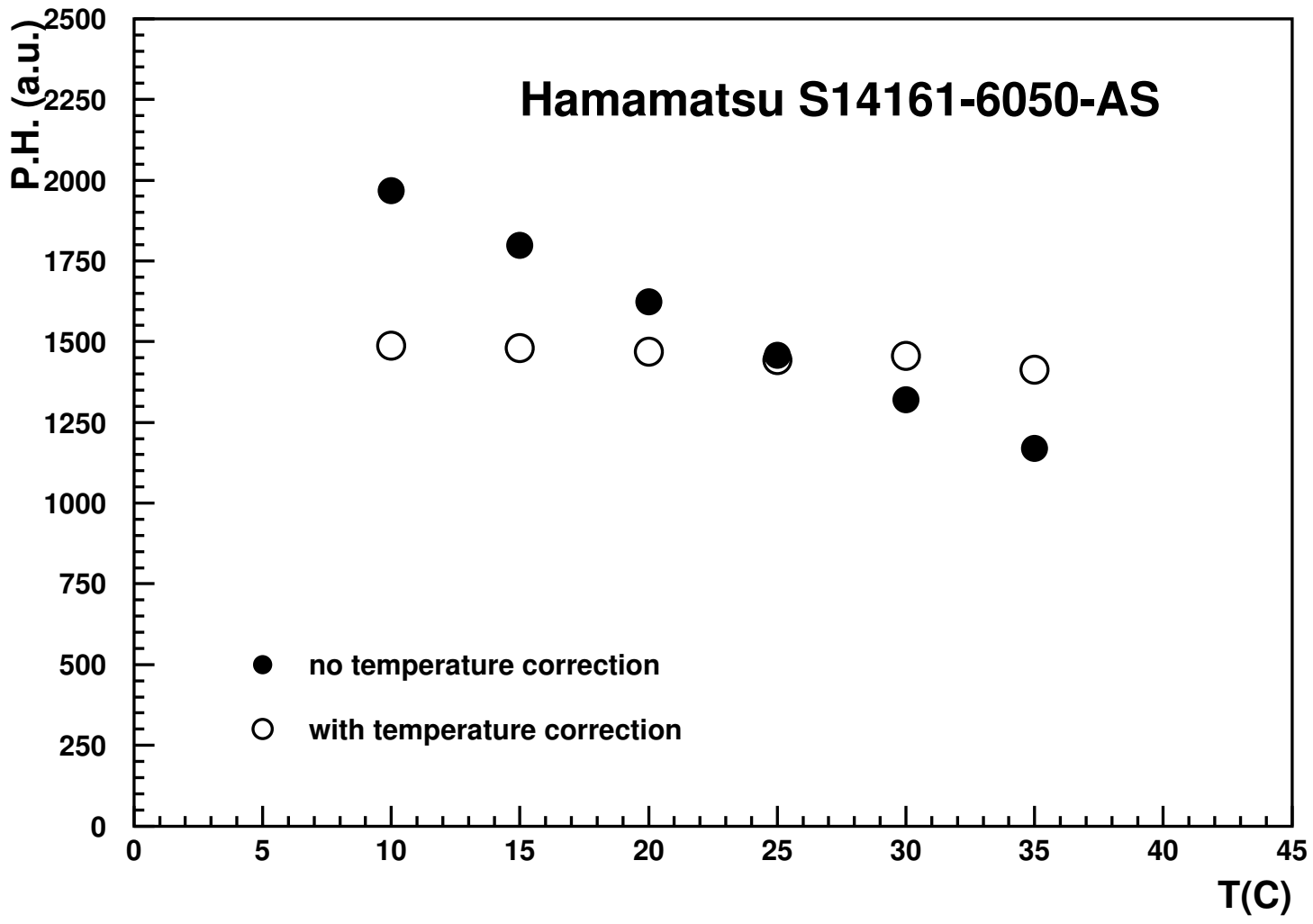

**Figure 2.** Photopeak position (a.u.) versus temperature T for a typical Ce:LaBr$_3$ crystal exposed to a Cs$^{137}$ source with and without online correction for the gain drift.

The custom NIM module has up to eight channels and an interface with the control PC based on the I2C protocol via an FDTI USB-I2C module or an Arduino one. Our approach is based on commercially available power supply modules (A7585D from CAEN), while other methods are based on ad hoc custom solutions, as proposed in references [34,35].

*Ganging of SiPM in One SiPM Array*

The SiPMs used in a SiPM array may be connected in different ways, depending on requirements such as speed, signal-over-noise ratio (S/N), and granularity. The different options are shown in Figure 3. In parallel ganging, the increased capacitance implies slow risetimes and long falltimes. In addition, there is the need to group SiPM with the same operating voltage $V_{op}$. In series ganging, instead, the charge/amplitude is reduced. This means faster signals but requires higher bias voltages: a factor $\times$ N with N number of single SiPMs. In hybrid ganging, single SiPMs are connected in series for signal and in parallel for bias, with decoupling capacitors in between, as originally developed for the MEG II upgrade [36]. A common bias voltage is used. The layout of the different ganging configurations is shown in Figure 3. For the waveforms with different ganging schemes, we have that

- Time constant: Series ∼ Hybrid < Parallel
- Pulse height: Series ∼ Hybrid > Parallel

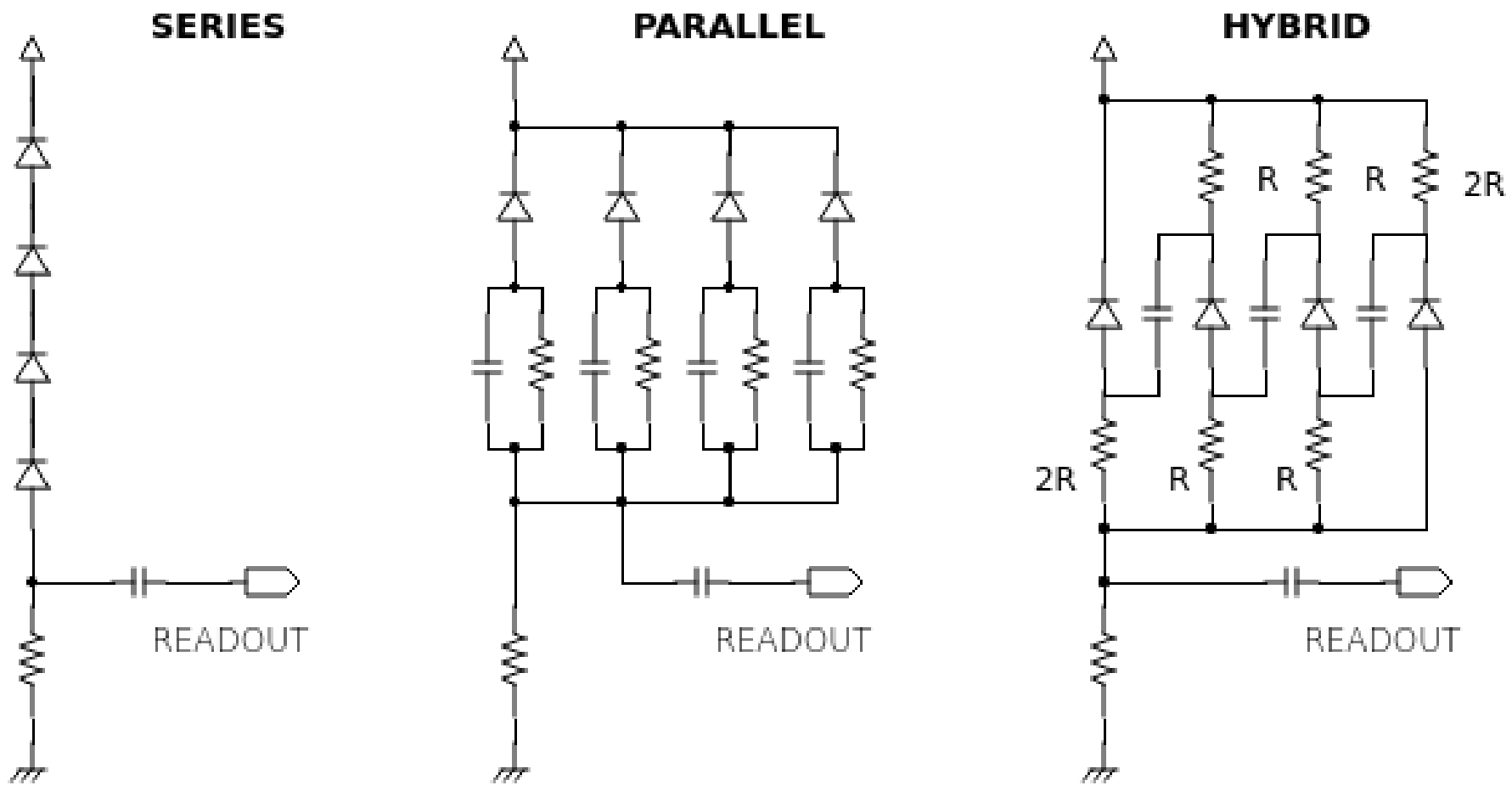

**Figure 3.** Layout of different ganging schemes for SiPMs: series ganging, parallel ganging, hybrid ganging from left to right.

The layouts of the circuits realized for the standard parallel ganging and the hybrid ganging in our laboratory tests are shown in Figures 4 and 5. From a custom design developed at the INFN Pavia electronics laboratory, based on what was suggested in references [37,38], they were realized using Phoenix srl, Ivrea. The 1″ (1/2″) SiPM arrays are mounted on the PCBs via two (one) SAMTEC multipin connectors.

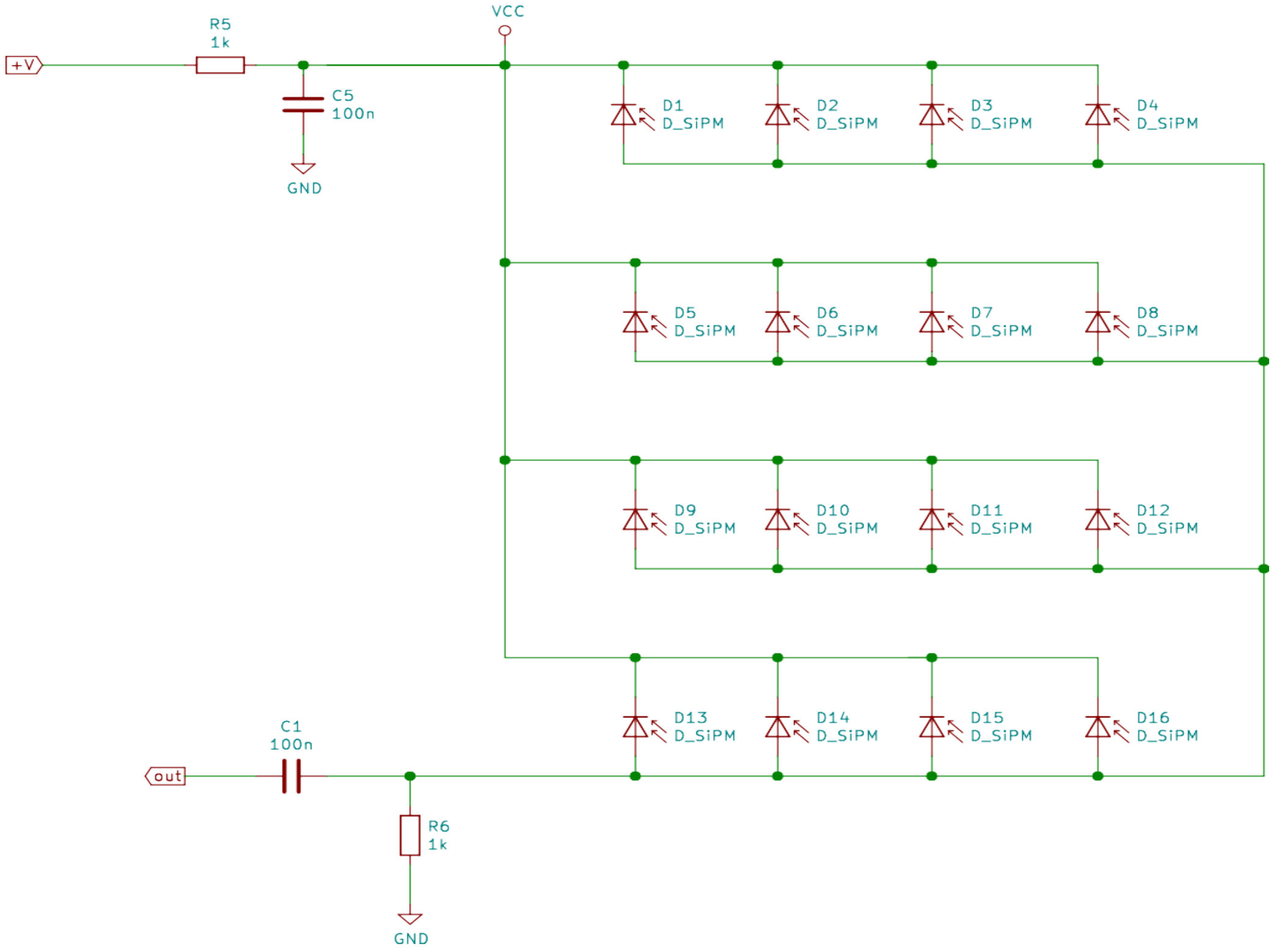

**Figure 4.** PCB circuit for 1″ crystals SiPM array mounting, with parallel ganging.

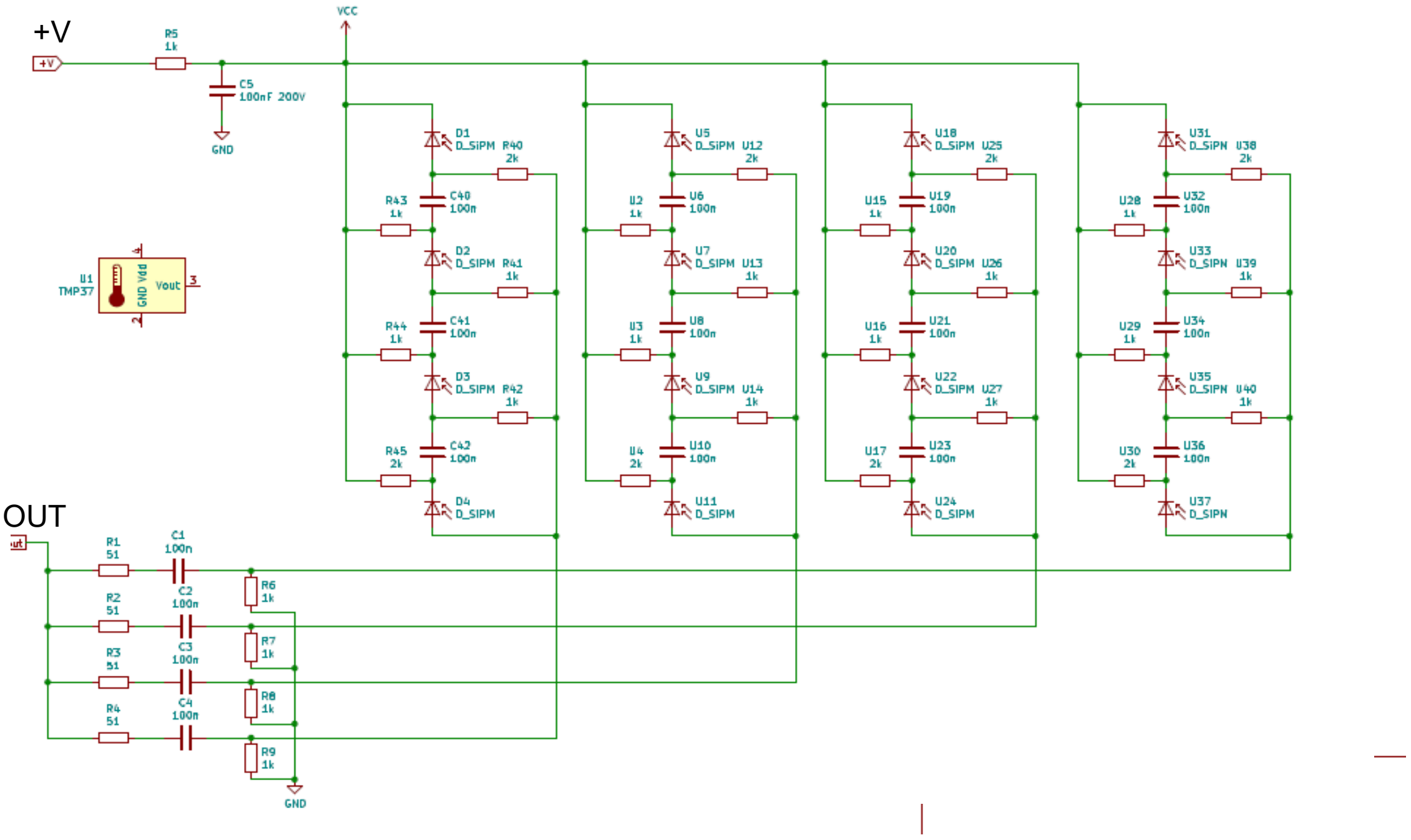

**Figure 5.** PCB circuit for 1″ crystals SiPM array mounting, with hybrid ganging.

The 4-1 Nuclear Instruments circuit is based on the idea of dividing the 1″ square SiPM array into four sub-arrays to reduce the capacitances involved and treat the zero pole compensation and amplification separately in each one. As shown in Figure 6, in the initial stage (stage 1), the signal from each sub-array has a pole-zero compensation, followed by amplification via Texas Instruments OPA695 amplifiers. Signals are then added in stage 2. The following stages realize an AC coupling (to cancel offsets) and invert the output signal.

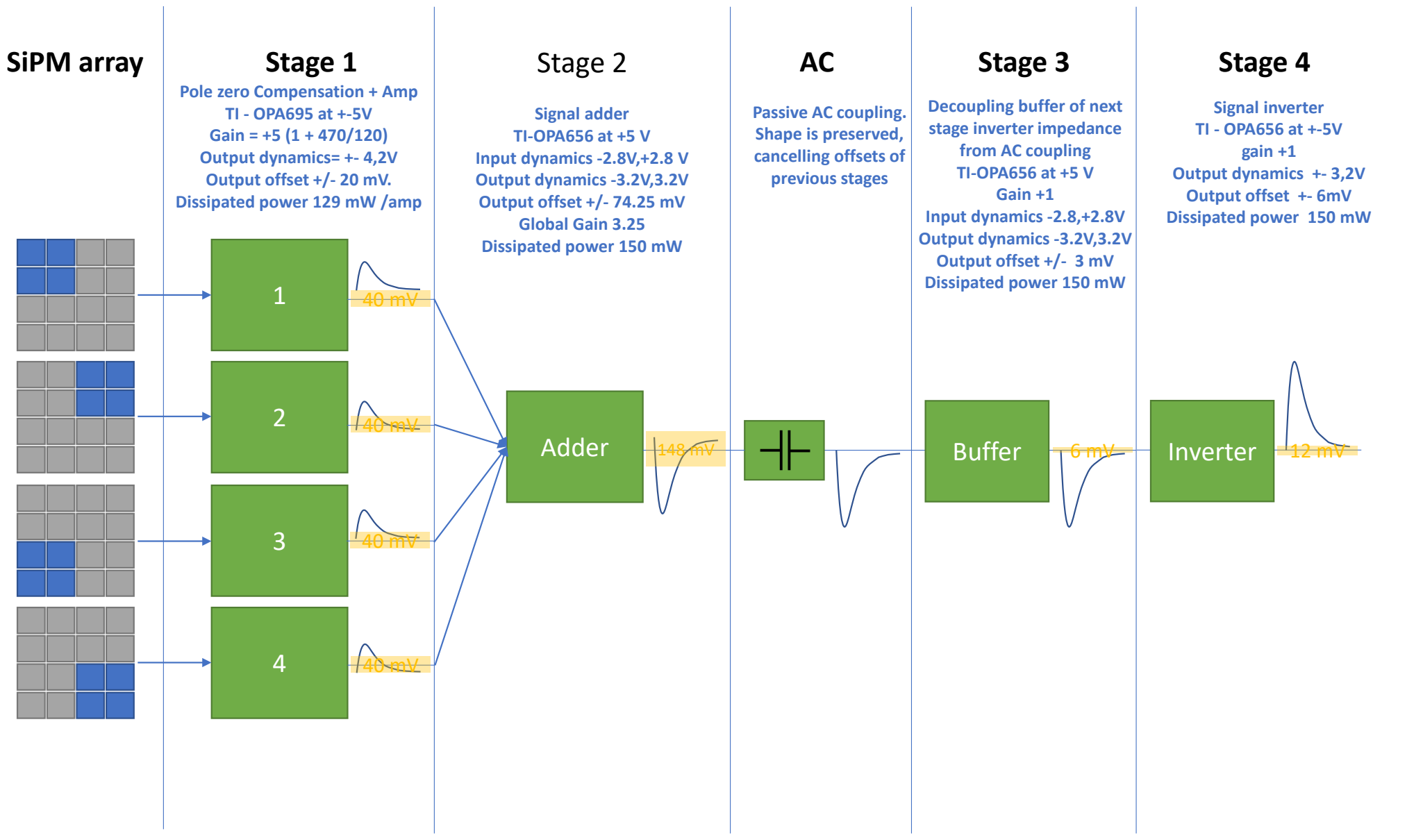

**Figure 6.** Behavior of the various stages of the 4–1 Nuclear Instruments circuit.

The schematics of the main circuit components are shown in Figure 7.

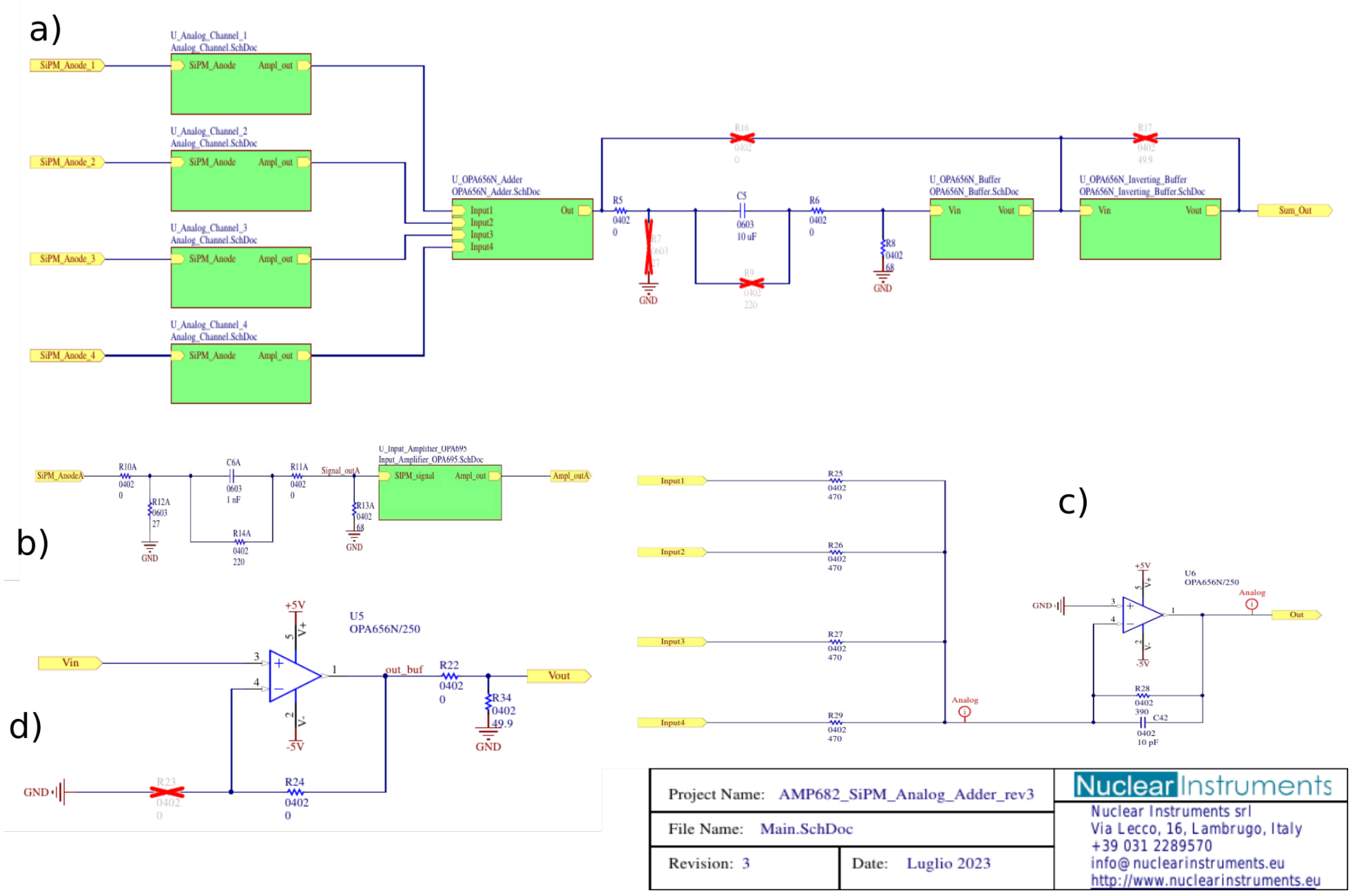

**Figure 7.** Schematics of the Nuclear Instruments 4-1 PCB circuit: **a**) the layout of the processing chain, **b**) the layout of the first amplification stage, **c**) the layout of the adder, and **d**) the layout of the buffer. An inverting amplifier is the last stage.

Images of the realized PCBs are shown in Figure 8. To use pre-existing mechanics, there were severe constraints on the final size of the PCB that had to fit inside a maximum size of $34 \times 34$ mm$^2$, thus requiring a compact design.

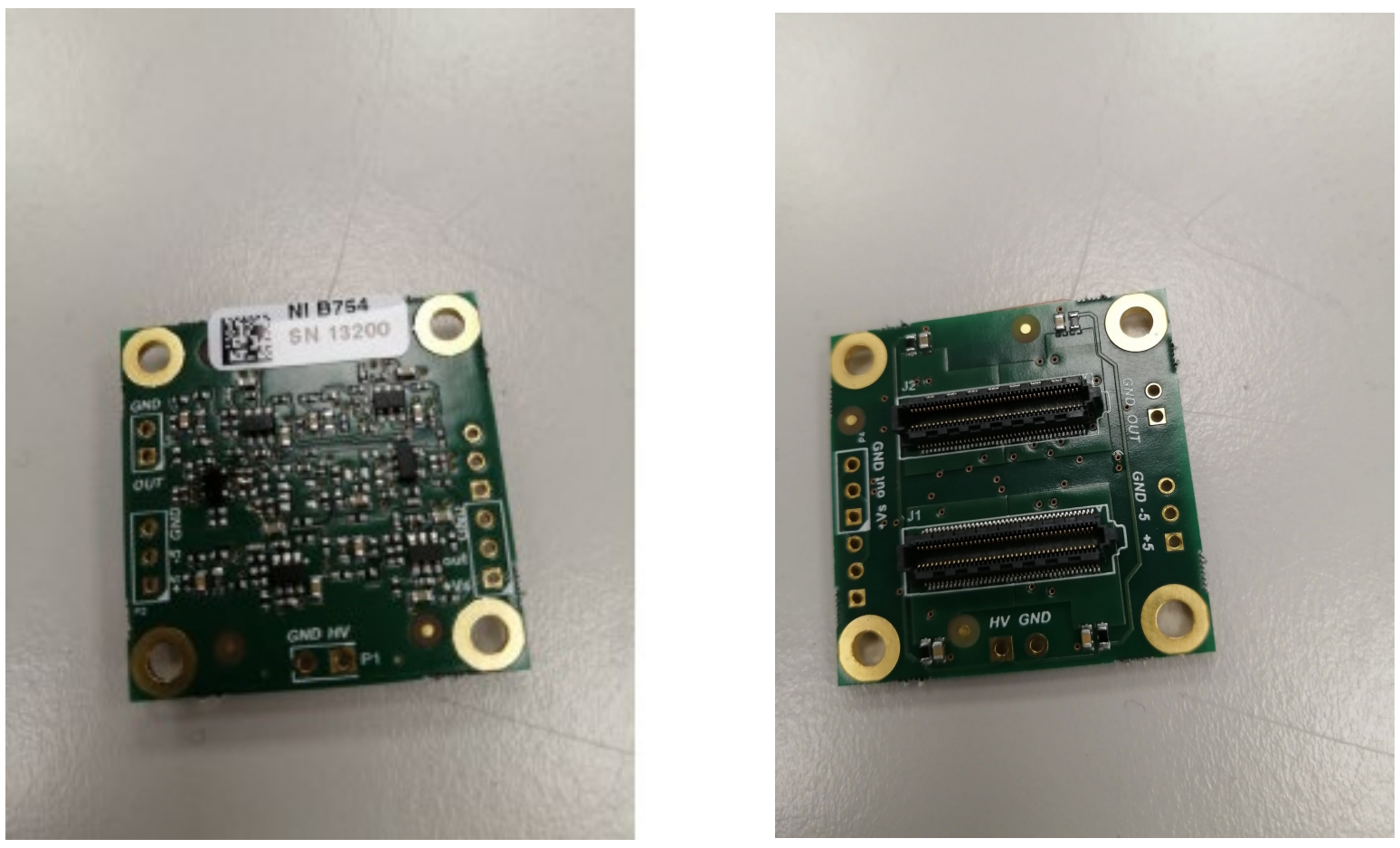

**Figure 8.** Bottom and top and pictures of the 4-1 PCB realized using Nuclear Instruments.

With this solution, there is a temperature increase of about 5–7 °C due to the dissipated power from the op-AMP used at the PCB level($\sim$1 W). To comply with it, a heat dissipator was put in thermal contact with the back of the PCB via a gap filler pad (see a in the left panel of Figure 1). Detectors' output signal and SiPM array powering are made via two coax single cables, while the TMP37 thermistor signal and the $\pm$5 V powering of the used OPA695 amplifier is via a 4-wire cable.

## 3. Results

Laboratory tests were performed with exempt sources from Spectrum Techniques ($Cd^{109}$, $Co^{57}$, $Ba^{133}$, $Na^{22}$, $Cs^{137}$, $Mn^{57}$) covering a range of X-rays' energies from 88 keV to 1274.5 keV. Detectors' signals were fed directly into a CAEN V1730 FADC and the data acquisition was via a custom DAQ developed for the FAMU experiment [39]. The produced n-tuples were analyzed using PAW [40] or ROOT [41] programs. All tests are performed inside a climatic chamber Memmert IPV-30 at a fixed temperature. For timing measurements, signals were visualized on a 1 GHz Lecroy scope.

### *3.1. Performances for a Typical Detector*

Timing and energy resolution results for a typical detector are shown in Table 1 and Figure 9, using different types of ganging for the SiPM arrays' cells.

**Table 1.** Results for a typical 1″ detector with different ganging.

| | $V_{op}$ (V) | Risetime (ns) | Falltime (ns) | Resolution % $Co^{57}$ | Resolution % $Cs^{137}$ |
|---|---|---|---|---|---|
| parallel | 40.82 | 68.9 ± 7.8 | 293.3 ± 43.4 | 7.78 | 2.96 |
| hybrid | 41.82 | 16.1 ± 2.4 | 176.8 ± 29.0 | 9.58 | 6.08 |
| 0-pole: 2nF | 43.02 | 58.2 ± 15.6 | 123.4 ± 21.7 | - | 2.99 |
| NI 4-1 circuit | 40.82 | 28.4 ± 4.5 | 140.6 ± 21.7 | 7.89 | 2.98 |

With both the hybrid ganging solution and zero pole suppression + increased SiPM overvoltage: +2.2V to compensate for signal reduction, a good timing may be achieved. Unfortunately, a good FWHM energy resolution may be obtained only with the second solution at the expense of a possible increase in the SiPMs' dark count rate (for more details, see reference [33]). An optimal compromise is instead obtained with the new 4-1 Nuclear Instruments solution, where at nominal $V_{op}$, the risetime (and falltime) of the signal is reduced by a factor ~2, with respect to parallel ganging, while keeping the FWHM energy resolution at the same level.

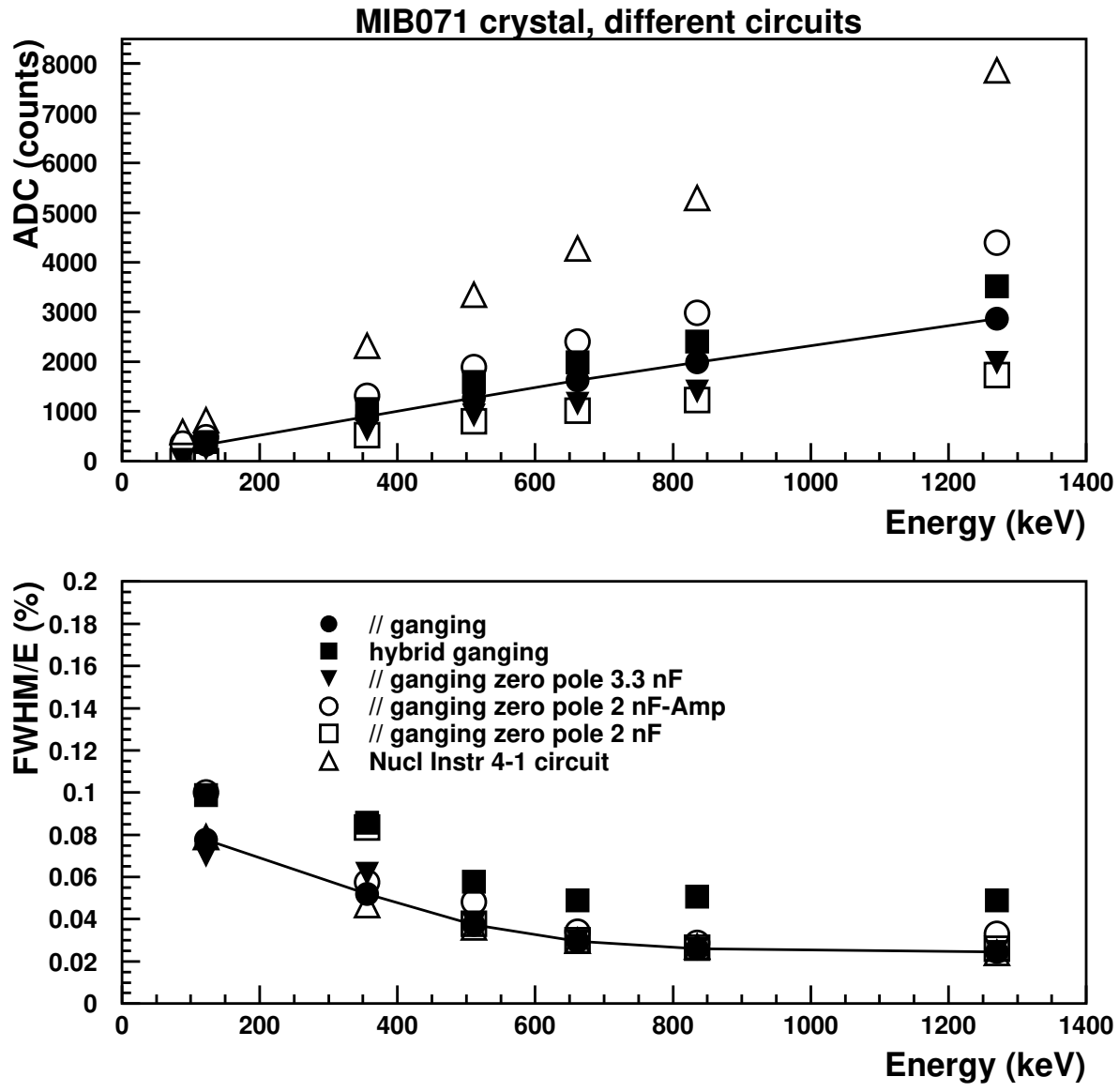

**Figure 9.** Linearity and FWHM energy resolution for a typical 1″ LaBr3:Ce crystal with different readout circuits. The line connects results with the standard parallel ganging of the SiPM cells for the SiPM array under test.

### *3.2. Results for the Whole Sample of Detectors Equipped with NI 4-1 PCB*

For the sample of 1″ round Ce:$LaBr_3$ detectors used in the FAMU experiment X-ray detector system, Figure 10 shows linearity and FWHM energy resolution as measured inside a Memmert IPV-30 climatic chamber at 20 °C (this temperature reflects the average temperature measured at Port 1 of RIKEN-RAL, where the FAMU experiment is installed) at INFN Milano Bicocca.

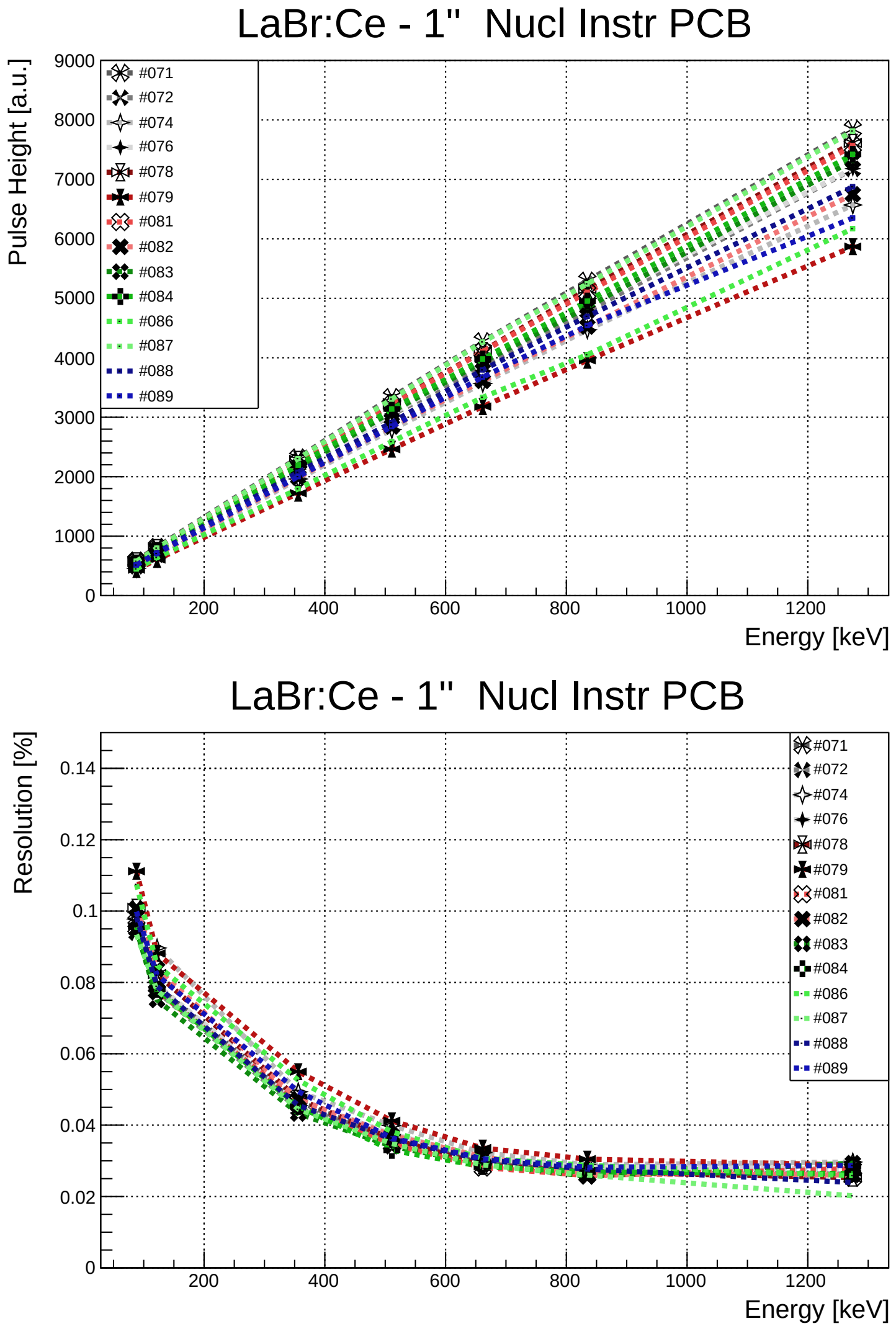

**Figure 10.** Linearity (**top** panel) and FWHM energy resolution (**bottom**) for a sample of 1″ round Ce:$LaBr_3$ crystals read by Hamamatsu S14161-6050AS-04 SiPM arrays. The detectors use a 4-1 PCB from Nuclear Instruments.

The timing properties for the same sample of detectors are shown instead in Figure 11 with a standard parallel ganging (top panels) and the new 4-1 solution (bottom panels). A 10–90 % risetime and falltime are reported, as measured on a 1 GHz Lecroy Wavesurfer 104MXs scope. The increase in timing properties is evident.

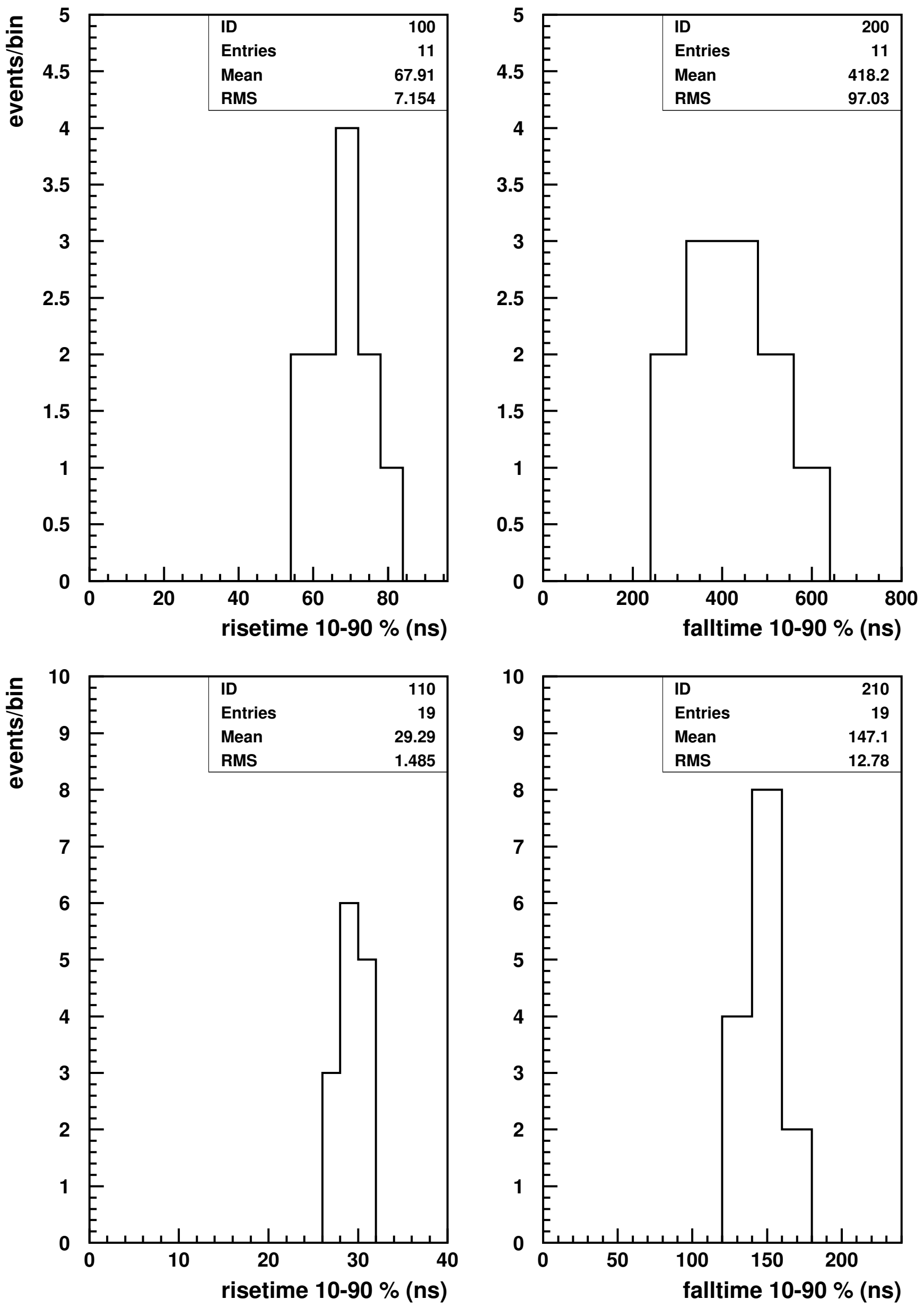

**Figure 11.** A 10–90 % risetime (falltime) for the case of standard parallel ganging (Nuclear Instruments 4-1 ganging) in the top (bottom) panels.

Several detectors were tested for the stability of response in time. They were put inside a Memmert IPV30 climatic chamber. Figure 12 shows a typical result. After the stabilization of the heat dissipation, the response measured at the $Cs^{137}$ photopeak is well within a $\pm 1\%$ band, around the average value, on a timescale of several hours.

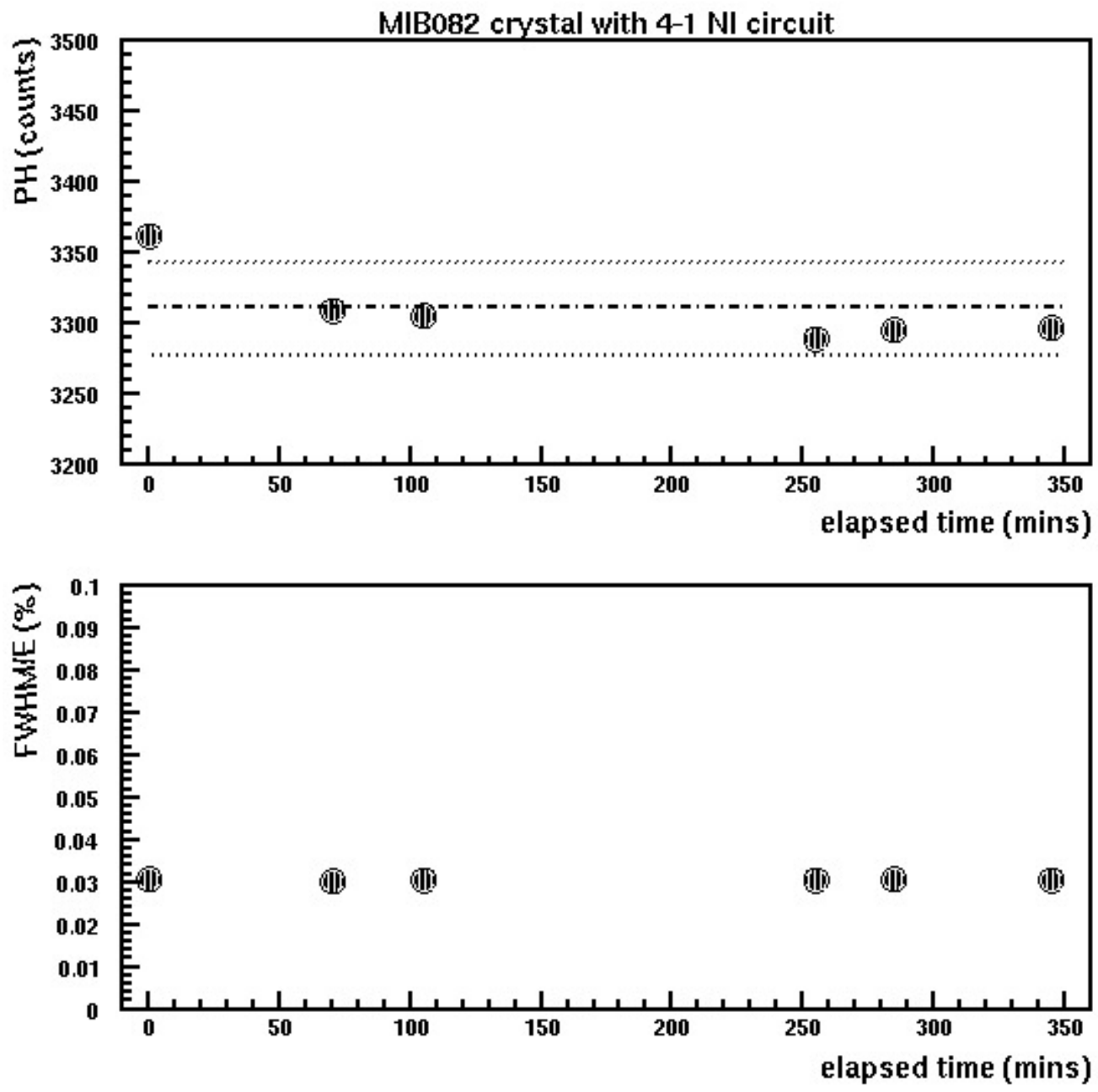

**Figure 12.** (**Top panel**) Recorded pulse height (P.H.) in a.u. and FWHM energy resolution for a typical detector versus elapsed time. The band represents a $\pm 1\%$ spread with respect to the average value. (**Bottom panel**) Same for the FWHM energy resolution.

Ten 1″ detectors and twelve 1/2″ detectors [12] are presently installed in the FAMU experiment, mounted on one upstream and one downstream crown, as shown in Figure 13. In between a central crown, detectors are held with a PMT readout, under repair for the breakdown of an electronic PCB. They are replaced now by six old detectors with a PMT readout [42] and six spare 1″ detectors with a SiPM array readout.

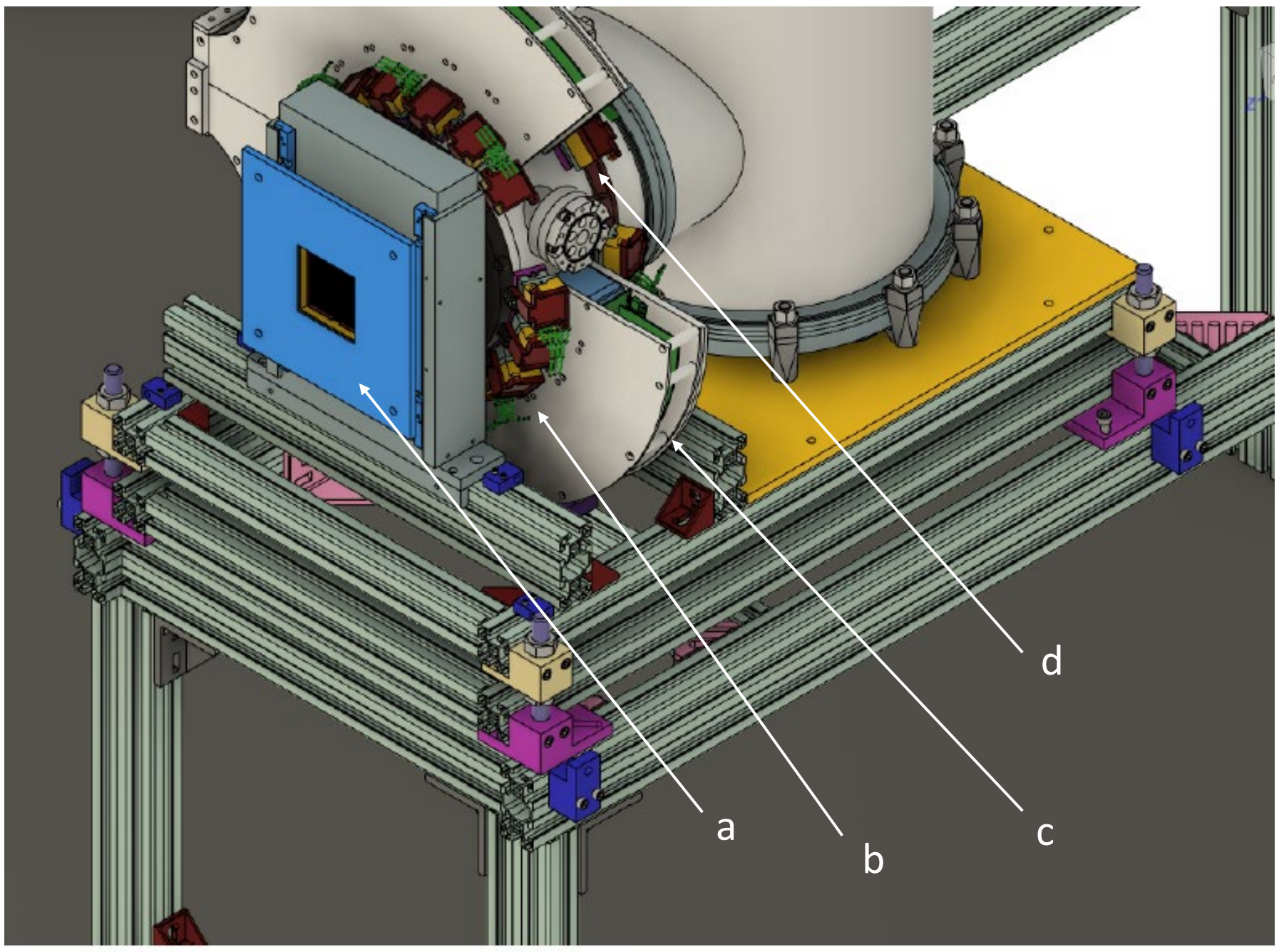

**Figure 13.** Image of the FAMU detector with (a) the beam hodoscope in front of a Pb collimator, (b) the upstream crown of 1″ $LaBr_3$:Ce detectors; (c) the central crown of detectors with a PMT readout, presently under repair; (d) the downstream crown of 1/2″ $LaBr_3$:Ce detectors.

Data were taken since March 2023 in Port 1 at RIKEN RAL and a preliminary analysis is under way. No major issues have been encountered up to now.

## 4. Discussion

Our results for FWHM energy resolution, obtained with 1″ detectors of reduced length (0.5″), with either the standard parallel ganging or the innovative 4-1 solution, compare well with the best results obtained with either a PMT or a SiPM readout. The timing properties of the signal pulse may be improved with a hybrid ganging, at the cost of a deteriorated FWHM energy resolution or with a zero pole circuit with the parallel ganging, at the cost of increasing the operating voltage, as shown in reference [33]. With the 4-1 innovative circuit from Nuclear Instruments that divide the readout of a 1″ SiPM array into four parts, a good compromise in the optimization of energy resolution and pulse timing is obtained. The major drawback of this solution is the increase in the dissipated heat due to the introduction of seven Texas Instruments OPA695 amplifiers per PCB. The total power dissipation is around 1 W. As the working environment is kept at a constant temperature (20 °C) using air-conditioning, a simple passive heat dissipation is enough for all detectors for proper operations.

The main characteristics of the Ce:$LaBr_3$ detectors used in the FAMU experiment, as measured in laboratory, are reviewed in Table 2.

**Table 2.** FWHM energy resolution and timing characteristics of the Ce:$LaBr_3$ detectors used in the FAMU experiment at RIKEN-RAL .

| | **Risetime (ns)** | **Falltime (ns)** | **Resolution %** $Cs^{137}$ | **Resolution %** $Co^{57}$ |
|---|---|---|---|---|
| 1/2″ detectors | 42.8 ± 4.7 | 372.4 ± 17.4 | 3.27 ± 0.11 | 8.44 ± 0.63 |
| 1″ detectors | 29.3 ± 1.5 | 147.1 ± 12.8 | 3.01 ± 0.16 | 7.93 ± 0.38 |

The worse timing properties of the 1/2″ detectors, as compared to the 1″ ones, are mainly due to the adoption of a standard parallel ganging instead of the 4-1 solution from Nuclear Instruments and probably to a different Ce concentration, as they come from a different producer.

## 5. Conclusions

Good FWHM energy resolution is obtained with 1″ Ce:$LaBr_3$ crystals read by the Hamamatsu S14161-6050AS-04 SiPM arrays. Resolutions better than 3%(8%) are obtained at the $Cs^{137}$ ($Co^{57}$) peak. The use of the innovative 4-1 circuit from Nuclear Instruments allowed a factor-two reduction in signal risetime (falltime) with respect to the conventional solution with parallel ganging. Solutions based on hybrid ganging instead show a sensible deterioration of FWHM energy resolution and were thus discarded.

**Author Contributions:** M. Bonesini: research design, article writing, data analysis, detectors' laboratory tests; R. Bertoni data analysis, detectors' laboratoty tests; A. Abba electronics development; F. Caponio electronics development, detectors' laboratory tests; M. Prata electronics development; M. Rossella electronics development

**Funding:** Research funding from INFN.

**Data Availability Statement:** Data may be available at request

**Acknowledgments:** We would like to thank all members of the FAMU collaboration for help and friendly discussions, in particular Andrea Vacchi, Ludovico Tortora, E. Mocchiutti and Riccardo Rossini. We acknowledge also the help in the mounting of detectors and the realization of the test setup of Luca Pastori, Nuclear Instruments srl, Roberto Gaigher and Giancarlo Ceruti; INFN Milano Bicocca mechanics workshop; and of Maurizio Perego, INFN Milano Bicocca, for SPICE simulations of the electronic circuits used.

**Conflicts of Interest:** Authors declare no conflict of interest.